# Citations vs journal impact factor as proxy of quality: could the latter ever be preferable?[1]

*Giovanni Abramo*[a,b,*], *Ciriaco Andrea D'Angelo*[a] *and Flavia Di Costa*[a]

[a] Laboratory for Studies of Research and Technology Transfer
School of Engineering, Dept of Management
University of Rome "Tor Vergata"

[b] National Research Council of Italy

**Abstract**

In recent years bibliometricians have paid increasing attention to research evaluation methodological problems, among these being the choice of the most appropriate indicators for evaluating quality of scientific publications, and thus for evaluating the work of single scientists, research groups and entire organizations. Much literature has been devoted to analyzing the robustness of various indicators, and many works warn against the risks of using easily available and relatively simple proxies, such as journal impact factor. The present work continues this line of research, examining whether it is valid that the use of the impact factor should always be avoided in favour of citations, or whether the use of impact factor could be acceptable, even preferable, in certain circumstances. The evaluation was conducted by observing all scientific publications in the hard sciences by Italian universities, for the period 2004-2007. Performance sensitivity analyses were conducted with changing indicators of quality and years of observation.

# 1. Introduction

In recent years, many nations have set a higher priority on the evaluation of research activity in the public sphere. National evaluation exercises have become increasingly common, so that funds can be allocated on the basis of performance criteria, as well as to stimulate greater research productivity and, last but not least, to reduce asymmetric information between knowledge suppliers and users. However, it is clear that a tradeoff has emerged in these evaluation exercises between "administrative" needs, which favour the use of systems based on simple measures, and somewhat contrasting requests from researchers for transparent, exhaustive and equitable evaluations of the real scientific quality of their work (Aksnes and Rip, 2009). The new trend of greater use of bibliometric indicators to support peer-review assessments (informed peer-review), has induced bibliometricians to pay greater attention to describing and resolving methodological problems: among these being the choice of more appropriate indicators for evaluation of the quality or impact[2] of scientific publications and, consequently, the research activity of single scientists, research groups and entire institutions.

Bibliometrics permits measurement of an important performance indicator, productivity, which can not be evaluated by peer review, within acceptable limits of cost and time. Weighting output for its relative quality, it is also possible to arrive at productivity measures that embed the attribute of quality. In bibliometrics, quality can be measured using two fundamental proxies: the citations received for a publication and the impact factor of the journal of publication. Good sense and common agreement is that, in general, the first of these indicators is preferable, and there is a rich literature in favour of this assertion. The use of impact factor for evaluation purposes has been the object of numerous criticisms by bibliometric experts (Moed and Van Leeuwen, 1996, Weingart, 2004). As early as 1997, Seglen observed that the journal can not be representative of the article, since it is the citation rates of the articles that determine the journal impact factor, and not vice versa. Even though impact factor offers characteristics of robustness, comprehensibility and methodological reproducibility, and is more immediately and easily obtained than citations (Bordons et al., 2002), the literature points out a range of limits and biases (Glanzel and Moed, 2002) that call for caution and careful attention in its use: i) a bias in favour of those journals that publish very long articles, such as review journals, ii) the fact that average time for journal articles to go from publication to their peak in citations can easily exceed two years; iii) the fact that the impact factor is a mean value calculated on the basis of citations for the articles in a journal, which is typically a very skewed distribution, and thus is a value that does not appropriately describe the overall pattern. Other criticisms concern a lack of consistency in identifying "citable documents": the ISI excludes recognition of citations for certain document types (particularly letters and editorials) and which are in fact actually cited, thus contributing to the impact factor identified for journals (Moed and Van Leeuwen, 1995).

In the context of the continuing debate on these methodological issues, the authors wish to verify if indeed the use of the impact factor as a proxy of quality should always be avoided in favour of citations or if, in certain circumstances, its application can be as appropriate as or even preferable to the use of citations. Here, it should be noted that bibliometric evaluation exercises, as for all systems of outcome control, especially if intended to support the allocation of resources, must be based on observations of the most recent past. Yet it is well known that the reliability of citations in representing the true quality of a scientific article depends on the very factor of the lapse between the publication date of the article and the moment of observing the number of citations received. Further, the life cycles of citations and the peaks in the citation distribution curves vary from discipline to discipline. From an analysis of the distributions of citations, Garfield (1972) observed early on that age distributions of cited references vary significantly among disciplines. For example, in examining the time distribution of citations for the two research fields of biochemistry-molecular biology and mathematics, it can be observed that in the first field the peak of the distribution is at an average of two years from publication, while for the second field the peak in citations is reached an average of a year later (Moed, 2005).

---

[2] We will use the two as synonyms

It can thus be hypothesized that, for assessments referring to time periods very close to the date of conducting the actual exercise, and especially for certain disciplines, the impact factor could turn out to be as a useful surrogate of the effective quality of an article, a predictor that is not necessarily less reliable than citations. To demonstrate this we will focus on the scientific output in the hard sciences for all Italian universities. First we will verify the existence of a correlation between rankings based on two indicators of productivity, calculated for the same triennium of observation, with output weighted for quality respectively by impact factor and by citations. Second, assuming that the rankings based on citations observed in the "mature" phase of the lifecycle of publications provide a valid benchmark, we will compare the shift from this benchmark when rankings are obtained from two further distinct exercises: the first based on citations and the second based on impact factor, but with both conducted at little time from the period of observation.

In this work, the next section describes the dataset used and the methodological assumptions. Section three presents and comments on the results obtained from the analysis for the 2004-2006 triennium. Section four proceeds to an intertemporal comparison that examines the sensitivity, in terms of shifts in the rankings, to changes in the moment of observing the citations. The last section presents the pertinent conclusions from the work and the authors' comments.

## 2. Research pattern, dataset and indicators

The structure of the Italian university system is based on the classification of research staff into scientific disciplinary sectors (SDSs). Each scientist is officially assigned to one among 370 SDS, which are in turn grouped in 14 university disciplinary areas (UDAs). The SDSs represent the unit of observation: in our model for analysis of productivity, for each university active in each SDS, the dimension of research staff constitutes the input variable, while output is considered to be the international scientific publications by the scientists in the SDS. The procedure for the analysis consists of four steps:
  i. completion of a census of scientific articles authored by researchers on staff in Italian universities;
 ii. association of these articles with:

a. indicators of their quality, standardized with respect to the relative means of the ISI subject categories in which the articles are listed[3]
      b. the official SDSs to which the authors belong;
iii. aggregation of output by university and SDS and standardization with respect to number of staff, to obtain a measure of productivity at the level of SDS;
iv. aggregation of the data at the level of UDAs, through the standardization and weighting of the data in reference to the SDS that compose them, to obtain robust rankings with respect to the intrinsic heterogeneity of the SDSs[4].

Given the objectives of the study, step ii.a involves two operations for each article: one which considers an indicator of quality linked to citations, and one which considers the impact factor of the journal. This provides two distinct ranking lists, suitable for comparisons.

**2.1. Dataset**

The data on output used in the study are obtained from the Observatory on Public Research in Italy, a bibliometric database developed by the authors and based on the Web of Science (WoS), which provides a census of scientific production of all research organizations situated in Italy. Beginning from the WoS data, a procedure for address recognition and reconciliation was used to extract all publications in the hard sciences, authored by Italian universities. By developing a complex algorithm for the disambiguation of the identity of the authors, it is possible to precisely attribute each publication to the university scientists who wrote it (for details, see Abramo et al. 2008b). The possibility of attributing each publication to individual researchers, and thus to the SDSs to which they belong, permits the application of measures of productivity at various levels of aggregation. In this work, beginning from the level of SDS, and by applying successive operations of weighting and standardization, we calculate the values of aggregate productivity for each UDA and for each university. This study considers the 183 hard science SDSs that compose 8 UDAs: Mathematics and computer sciences, Physics, Chemistry, Earth sciences, Biology, Medicine, Agricultural and veterinary sciences, Industrial and information engineering[5]. The research staff considered is composed of the assistant, associate and full professors belonging to the SDSs examined. This represents a total of 34,163 scientists (averaged over the period under consideration) as identified on the basis of the CINECA database of the Italian Ministry of Research[6]. Table 1 presents the data on the staff and scientific production of Italian universities by UDA, for 2004-2006, which is our benchmark period.

---

[3] Thomson Reuters classifies each article indexed in Web of Science under a specific ISI subject category. For details see http://science.thomsonreuters.com/cgi-bin/jrnlst/jloptions.cgi?PC=D

[4] Data standardization serves to eliminate bias due to the different publication "fertility" of the various sectors within a single area, while data weighting takes account of the diverse presence of the SDSs, in terms of staff numbers, in each UDA (Abramo et al., 2008a)

[5] Civil engineering and architecture was excluded from the analysis because WoS listings are not sufficiently representative of research output in this area.

[6] http://cercauniversita.cineca.it/php5/docenti/cerca.php

| UDA | Number of SDS (total) | Universities | Research staff (Average) | Number of publications (Total) |
|---|---|---|---|---|
| Mathematics and computer sciences | 10 | 61 | 3,224 | 6,558 |
| Physics | 8 | 59 | 2,549 | 11,729 |
| Chemistry | 12 | 58 | 3,196 | 13,976 |
| Earth sciences | 12 | 48 | 1,278 | 2,335 |
| Biology | 19 | 64 | 5,087 | 15,641 |
| Medicine | 50 | 58 | 11,030 | 27,331 |
| Agricultural and veterinary sciences | 30 | 52 | 3,121 | 4,860 |
| Industrial and information engineering | 42 | 63 | 4,678 | 10,939 |
| Total | 183 | 71 | 34,163 | 93,369* |

*Table 1: Research staff and scientific output of Italian universities (2004-2006) per university disciplinary area (UDA).*
*There were 81,483 actual publications, however the data reported reflect multiple counts due to co-authorship by researchers belonging to SDSs in different UDAs.

The number of active universities varies according to the UDA considered, from a minimum of 48 for Earth sciences to a maximum of 64 for Biology, with a total of 71 universities considered overall. The number of SDSs also varies from area to area: from a minimum of eight in Physics to a maximum of 50 in Medicine. Medicine is also the largest area in terms of staff, with its 11,030 researchers representing almost one third of the total number in all UDAs. Earth sciences, with 1,278 staff, is the smallest of the UDAs considered.

Among other possible observations, one can note the diverse scientific "fertility" of the various UDAs.

## 2.2 Indicators

Two indicators are applied for each publication in the dataset. The first refers to the citations for the article ($QI_A$), observed as of 31/03/2008, and the second to the impact factor of the journal ($QI_J$). Since the distributions are typically very skewed in all fields, it was seen appropriate to use the percentile as a means of standardization for both indicators, calculated with respect to the distribution of each ISI subject category. The quality indexes for individual publications are thus defined as follows:

$QI_A$ = Article Quality Index, measured on a 0 – 100 percentile scale, according to the citation distribution of publications of the same year falling in the same ISI category. A value of 90 indicates that 90% of the articles of the same year falling in the same ISI category have a lower number of citations than the one under observation.

$QI_J$ = Journal Quality Index, measured on a 0 – 100 percentile scale according to the impact factor distribution of the journals publishing papers in the same ISI category. A value of 90 indicates that 90% of publications falling in the same category are on journals with lower impact factor than the one under observation.

On the basis of the affiliation and classification of the authors, the next step is the aggregation of the publications by university and SDS. The output indicator considered in the study is defined as "Scientific Strength". This is calculated using two different methods, being given by the sum of the publications authored by the researchers of the SDS of a university, with each weighted according to the Article Quality Index ($QI_A$) or Journal Quality Index ($QI_J$) of the publication itself. In formulae, per a generic SDS of a generic university:

$$SS_A = \sum_i (b_i \cdot QI_{A_i})$$

Where:
$SS_A$ = scientific strength based on citations of the article,
$b_i$ = 1, if the SDS under observation features at least one author among those for the publication $i$; 0, otherwise.

and $SS_J = \sum_i (b_i \cdot QI_{J_i})$

The same calculations are applied for $SS_J$ = scientific strength based on journal impact factor, with $QI_J$ replacing $QI_A$.

At this point it is possible to calculate the productivity ($P$) of an SDS as a ratio of Scientific Strength and the number of scientists ($Add$) present in the SDS; respectively:

$$P_A = \frac{SS_A}{Add} \text{ and } P_J = \frac{SS_J}{Add}$$

Our analysis will be conducted at the level of UDAs. To obtain a value for performance at this higher level, the data for the SDSs are aggregated through an operation of standardization and weighting. The objective is to limit the distortions typical of aggregate analysis that do not account for the varying fertility of the SDSs and of their varying representation, in terms of numbers of staff, within each UDA.

Thus, the productivity ($P_A$) of a generic UDA of a generic university is calculated as follows:

$$P_A = \sum_{s=1}^{n} \left( \frac{P_{As}}{P_{As}^*} \cdot \frac{Add_s}{Add} \right)$$

where:
$P_{As}$ = Productivity of SDS $s$
$P_{As}^*$ = Average productivity of national universities in SDS $s$
$Add_s$ = number of scientists in SDS $s$
$Add$ = number of scientists in the UDA
$n$ = number of SDSs in the UDA

The same calculations are applied to obtain $P_J$, productivity in terms of the journal quality index.

## 3. Correlation between productivity rankings based on citations and on impact factor

Still referring to the 2004-2006 triennium, the rankings for the universities active in each UDA were elaborated, according to their values of $P_J$ and $P_A$. The citations for $P_A$ were observed as of 31/03/2008.

The next step was to calculate the correlation coefficient for the rankings based on the two indicators. For greater robustness the analysis excluded those universities that had less than six scientists on staff in the UDA examined, averaged over the triennium. The results, presented in Table 2, show there is a strong correlation for all UDAs (values near to or greater than 0.90), with all values positive and statistically significant.

| UDA | Number of universities | Correlation |
|---|---|---|
| Mathematics and computer sciences | 53 | 0.936 |
| Physics | 49 | 0.962 |
| Chemistry | 45 | 0.968 |
| Earth sciences | 39 | 0.899 |
| Biology | 52 | 0.963 |
| Medicine | 43 | 0.982 |
| Agricultural and veterinary sciences | 31 | 0.954 |
| Industrial and information engineering | 47 | 0.914 |

*Table 2: Correlation index for rankings by $P_A$ and $P_J$; data 2004-2006.*

Further statistics concerning the comparisons between rankings are presented in Table 3. Calculations were made, for each UDA of each university, of the change in ranking by $P_A$ or $P_J$. The portion of universities affected by a difference in rank ranges from 67.3% of the total, for Physics, to 86.8% for Mathematics and computer sciences. Earth sciences shows the case of the highest absolute variation for a university (19 places) between the two rankings. The maximum shifts in rankings seen in the other UDAs are less, but still notable, particularly in Industrial and information engineering (14 places), Mathematics and computer sciences (13), Biology (13) and Physics (12). With all the UDA, the mean value of variation in rank is greater than the median, indicating that the distribution of the differences is asymmetric to the right. Mathematics and computer sciences is again the area with least convergence between the two rankings, in terms of mean (4.3) and median variation (4.0) in ranking, followed by Industrial and information engineering (4.2 and 3.0, respectively).

| UDA | Variation | Max | Average | Median |
|---|---|---|---|---|
| Mathematics and computer sciences | 46 out of 53 (86.8%) | 13 | 4.3 | 4.0 |
| Physics | 33 out of 49 (67.3%) | 12 | 2.5 | 1.0 |
| Chemistry | 34 out of 45 (75.6%) | 8 | 2.4 | 2.0 |
| Earth sciences | 31 out of 39 (79.5%) | 19 | 3.4 | 2.0 |
| Biology | 39 out of 52 (75.0%) | 13 | 2.6 | 1.5 |
| Medicine | 35 out of 43 (81.4%) | 8 | 1.6 | 1.0 |
| Agricultural and veterinary sciences | 25 out of 31 (80.6%) | 6 | 2.1 | 2.0 |
| Industrial and information engineering | 39 out of 47 (83.0%) | 14 | 4.2 | 3.0 |

*Table 3: Comparison of rankings by PA and PJ; 2004-2006 data.*

Table 4 presents a more detailed examination, with the case of the Biology area selected as an example. In the 2004-2006 triennium, this UDA had 52 universities with at least 6 scientists. The table shows the lists of rankings for $P_A$ and $P_J$ with calculation of the variation in ranking. The changes vary from the extremes of -13 for the University of Venice "Ca' Foscari" to +12 for the University of Udine. Thirteen universities show the same ranking under both indicators while 43 (83% of total) show shifts of between -4 and +4. An overall view of the range of variation and average values of rankings is presented in Figure 1. The trend of the data is for a substantial convergence of rankings but, in the cases of some universities, with oscillations that are notably more than marginal.

| Univ_ID | University name | Research staff | $P_A$ rank | $P_J$ rank | Variation |
|---|---|---|---|---|---|
| 13 | University of Venice "Ca' Foscari" | 16 | 17 | 30 | -13 |
| 62 | University of Bologna | 222 | 20 | 29 | -9 |
| 36 | University of Modena and Reggio Emilia | 100 | 16 | 23 | -7 |
| 50 | University of Turin | 174 | 14 | 20 | -6 |
| 26 | University of Brescia | 47 | 23 | 28 | -5 |
| 54 | University of Urbino "Carlo Bo" | 66 | 30 | 34 | -4 |
| 17 | University of Molise-Campobasso | 25 | 39 | 42 | -3 |
| 22 | University of Viterbo "Tuscia" | 47 | 24 | 27 | -3 |
| 41 | University of Palermo | 172 | 49 | 52 | -3 |
| 49 | University of Teramo | 6 | 5 | 8 | -3 |
| 31 | University of Genova | 133 | 11 | 13 | -2 |
| 34 | University of Milan | 356 | 9 | 11 | -2 |
| 47 | University of Sassari | 75 | 43 | 45 | -2 |
| 55 | University of Verona | 58 | 12 | 14 | -2 |
| 9 | Scuola Normale Superiore in Pisa | 10 | 3 | 4 | -1 |
| 24 | University of Bari | 179 | 31 | 32 | -1 |
| 30 | University of Foggia | 12 | 48 | 49 | -1 |
| 44 | University of Perugia | 120 | 45 | 46 | -1 |
| 11 | Second University of Naples | 103 | 38 | 38 | 0 |
| 15 | Sacred Heart Catholic University | 64 | 36 | 36 | 0 |
| 19 | University of Lecce "Salento" | 46 | 15 | 15 | 0 |
| 27 | University of Cagliari | 127 | 41 | 41 | 0 |
| 33 | University of Messina | 127 | 37 | 37 | 0 |
| 43 | University of Pavia | 170 | 25 | 25 | 0 |
| 45 | University of Rome "La Sapienza" | 369 | 35 | 35 | 0 |
| 60 | University of Calabria | 56 | 40 | 40 | 0 |
| 61 | University of Varese "Insubria" | 61 | 6 | 6 | 0 |
| 64 | University of Catania | 145 | 47 | 47 | 0 |
| 65 | University of Ferrara | 107 | 7 | 7 | 0 |
| 68 | Polytechnic University of Ancona | 62 | 2 | 2 | 0 |
| 69 | University of Milan "Vita-Salute San Raffaele" | 13 | 1 | 1 | 0 |
| 2 | University of Rome "Foro Italico" | 6 | 51 | 50 | 1 |
| 18 | University of Eastern Piedmont "A. Avogadro" | 42 | 13 | 12 | 1 |
| 35 | University of Milan "Bicocca" | 69 | 34 | 33 | 1 |
| 37 | University of Naples "Federico II" | 267 | 32 | 31 | 1 |
| 39 | University of Naples "Parthenope" | 7 | 52 | 51 | 1 |
| 40 | University of Padua | 198 | 10 | 9 | 1 |
| 46 | University of Rome "Tor Vergata" | 155 | 27 | 26 | 1 |
| 57 | University of Catanzaro "Magna Grecia" | 30 | 44 | 43 | 1 |
| 59 | University of "Roma Tre" | 31 | 4 | 3 | 1 |
| 21 | University of Basilicata | 7 | 50 | 48 | 2 |
| 29 | University of Florence | 174 | 18 | 16 | 2 |
| 63 | University of Camerino | 68 | 46 | 44 | 2 |
| 23 | University of L'Aquila | 81 | 42 | 39 | 3 |
| 42 | University of Parma | 136 | 22 | 19 | 3 |
| 66 | University of Pisa | 168 | 21 | 18 | 3 |
| 67 | University of Salerno | 21 | 8 | 5 | 3 |
| 20 | University of Benevento "Sannio" | 14 | 28 | 24 | 4 |
| 52 | University of Trieste | 93 | 29 | 22 | 7 |
| 48 | University of Siena | 132 | 19 | 10 | 9 |
| 56 | University of Chieti "Gabriele D'Annunzio" | 50 | 26 | 17 | 9 |
| 53 | University of Udine | 41 | 33 | 21 | 12 |

*Table 4: Comparison between rankings by $P_A$ and $P_J$ for Italian universities in the Biology UDA; 2004-2006 data.*

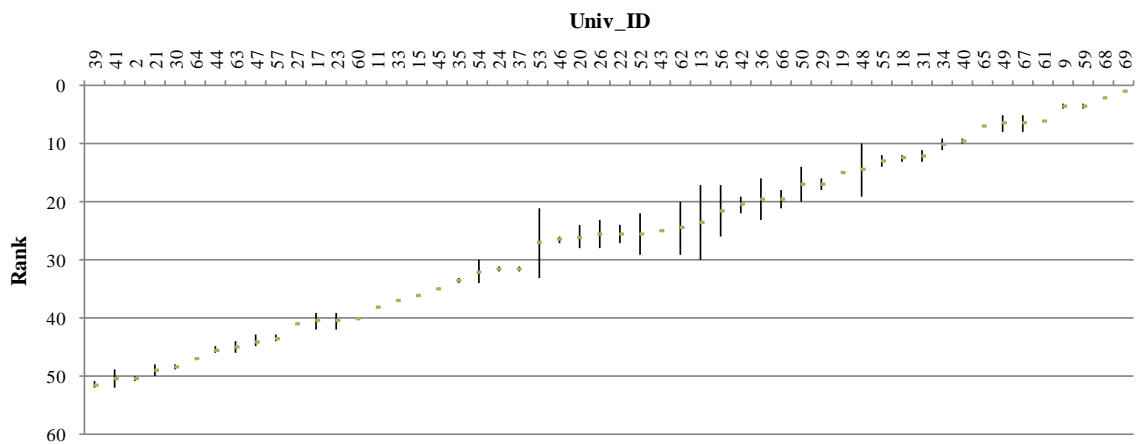

*Figure 1: Distribution of rankings (average and range of variation for $P_A$ and $P_J$) for Italian universities active in the Biology UDA.*

## 4. Inter-temporal analysis: sensitivity of productivity rankings to the date of observing quality indicators

In the previous section we have shown how, for the 2004-2006 triennium, there is a strong correlation between productivity rankings for Italian universities when the impact factor is used in place of citations as a proxy of quality for individual publications. There is also an occurrence of a number of outliers, universities that place in significantly different positions under the two rankings. In this section we will repeat the analysis for the single year of 2007, to examine whether this substantial "convergence" of the two rankings is still present when considering a period of observation that is very close to the date of conducting the evaluation exercise. We will continue to use the date of 31/03/2008 for observation of citations of all publications in the dataset. Table 5 presents the results of the comparison. It is readily apparent that the convergence of the two rankings is sharply reduced: the correlation coefficient falls notably in all UDAs and the portion of universities that change position in the two rankings is greater than 90%. In Mathematics and computer sciences, the correlation index for the $P_A$ and $P_J$ rankings falls to 0.467 (from 0.936 observed for the 2004-2006 triennium). Other statistics also indicate substantial differences between the two rankings. The average value of the changes when ranked by $P_A$ and $P_J$ varies from 4.3 for Agricultural and veterinary sciences to 11.5 for Mathematics and computer sciences, while for the 2004-2006 analysis these values were respectively 2.1 and 4.3. The median of the differences in ranking is also much greater than that registered for 2004-2006. The maximum change also results as truly notable: in Mathematics, for example, the University of Eastern Piedmont "A. Avogadro" drops from fifth position in rankings by $P_A$ to 52nd position in those by $P_J$. In Chemistry, for the 2004-2006 analysis, the maximum change in rankings was 8, while for 2007 the maximum shift increases to 39, with the University of Trento moving from 42nd position under $P_A$ to third place under the $P_J$ rankings. Among all the UDAs, those that show the most limited change in rankings are Medicine and Agricultural and veterinary sciences. In all other UDAs, the choice of the quality indicator for evaluation of publications results as truly decisive in result, giving list rankings that are very different. The UDA where this is most remarkable and problematic is clearly Mathematics and computer sciences.

|  | 2007 | | | | | 2004-2006 | | | | |
| --- | --- | --- | --- | --- | --- | --- | --- | --- | --- | --- |
| UDA | % Var | Max | Ave | Median | Corr | % Var | Max | Ave | Median | Corr |
| Mathematics and computer sciences | 96.4 | 47 | 11.5 | 8 | 0.467 | 86.8% | 13 | 4.3 | 4.0 | 0.936 |
| Physics | 89.8 | 37 | 8.2 | 5 | 0.660 | 67.3% | 12 | 2.5 | 1.0 | 0.962 |
| Chemistry | 93.8 | 39 | 8.1 | 4.5 | 0.621 | 75.6% | 8 | 2.4 | 2.0 | 0.968 |
| Earth sciences | 94.9 | 26 | 8.1 | 7 | 0.554 | 79.5% | 19 | 3.4 | 2.0 | 0.899 |
| Biology | 90.6 | 34 | 7.3 | 5 | 0.777 | 75.0% | 13 | 2.6 | 1.5 | 0.963 |
| Medicine | 91.3 | 19 | 5.2 | 3 | 0.849 | 81.4% | 8 | 1.6 | 1.0 | 0.982 |
| Agricultural and veterinary sciences | 90.3 | 20 | 4.3 | 3 | 0.783 | 80.6% | 6 | 2.1 | 2.0 | 0.954 |
| Industrial and information engineering | 97.9 | 32 | 9.1 | 6.5 | 0.617 | 83.0% | 14 | 4.2 | 3.0 | 0.914 |

*Table 5: Comparison between rankings by $P_A$ and $P_J$, for periods 2004-2006 and 2007.*

The analysis completed here shows that correlation between rankings for productivity obtained by weighting output for citations and for impact factor tends to diminish with lessening of the time elapsed between the date for observing the citations and the closing date for the time period subject to analysis. We were already aware that citations lose reliability as a proxy of article quality with lessening of time between date of observation of the citations and date of publication of the article. However, the results presented in Table 5 do not permit arrival at definitive conclusions concerning the greater or lesser reliability of either of the two proxies. This requires another analysis based on a valid benchmark, which would be the productivity ranking based on the citations observed at the "mature" phase of the publication life cycle. In our case, we have considered the productivity ranking referring to the output for the single year of 2004, weighted for citations observed as of 31/03/2008, which is at least 39 months after the publication of the articles. We repeat the analysis, this time for 2004 and in this case considering, as proxies for output quality, first the citations observed as of 01/01/2006 and next the impact factor of the journals as listed in the Journal Citation Report™ 2006. The question we intend to answer is: which of the two analyses provides the rankings that are closer to the benchmark. The response is presented in Table 6. For each UDA, the first five columns of figures present statistics concerning the variations in ranking between the benchmark and the analysis conducted using the number of citations observed as of 01/01/2006, and the next five columns present the same statistics for the analysis conducted using the impact factor.

|  | $P_A$ (citations 2006) vs $P_A$ (citations 2008) | | | | | $P_J$ (JCR 2006) vs $P_A$ (citations 2008) | | | | |
| --- | --- | --- | --- | --- | --- | --- | --- | --- | --- | --- |
| UDA* | % Var | Max | Ave | Median | Corr | % Var | Max | Ave | Median | Corr |
| Mathematics and computer sciences | 84.9% | 21 | 5.3 | 4 | 0.886 | 86.8% | 20 | 4.3 | 3 | 0.916 |
| Physics | 75.0% | 19 | 2.7 | 1 | 0.949 | 83.3% | 12 | 2.9 | 2 | 0.957 |
| Chemistry | 77.8% | 14 | 2.5 | 2 | 0.956 | 93.3% | 18 | 3.8 | 2 | 0.915 |
| Earth sciences | 87.2% | 18 | 4.3 | 3 | 0.863 | 94.9% | 14 | 3.5 | 2 | 0.911 |
| Biology | 88.0% | 21 | 3.8 | 3 | 0.921 | 84.0% | 22 | 3.4 | 2 | 0.935 |
| Medicine | 78.0% | 7 | 1.5 | 1 | 0.984 | 78.0% | 7 | 2 | 2 | 0.976 |
| Agricultural and veterinary sciences | 77.4% | 12 | 3.1 | 2 | 0.874 | 83.9% | 14 | 4.6 | 3 | 0.767 |
| Industrial and information engineering | 97.8% | 43 | 8.7 | 2 | 0.570 | 95.6% | 43 | 8.5 | 2 | 0.523 |

*Table 6: Comparison between productivity rankings derived from observations made as of 2006 and 2008, for bibliometric data concerning 2004.*
*\* Observations exclude UDAs with an average of less than 6 scientists on staff over the triennium.*

The data indicate that the ranking based on citations is definitely closer to the benchmark in only two areas: Chemistry and Agricultural and veterinary sciences. For the specific case of Chemistry, apart from median variation, the statistics all indicate that citations offer greater reliability than impact factor in approximating the benchmark: correlation between the benchmark and the ranking based on impact factor is 0.915 compared to the 0.956 between the benchmark and the ranking based on citations; the average change in rank is more limited when considering citations (2.5) than when considering impact factor (3.8); the same also holds for maximum variation (14 compared to 18).

In the areas of Mathematics and computer sciences, Earth sciences, and Biology, it seems that the use of the impact factor for weighting output provides productivity rankings that are definitely closer to the benchmark. In Mathematics and computer sciences the correlation with the benchmark ranking is 0.916 using impact factor, compared to 0.886 using citations; the average variation in rank compared to the benchmark is 5.3 when using citations, but decreases to 4.3 if impact factor is used; the same situation occurs for the median of change in rankings, which declines from 4 to 3 when impact factor is used in place of citations.

In Physics, Medicine and Industrial and information engineering it is difficult to indicate which of the two proxies of quality provides rankings of productivity that are closer to the benchmark.

## 5. Conclusions

The proposed study is a further contribution to an area of methodological research concerning the suitability of impact factor as a proxy of quality for scientific publications. Numerous and authoritative works in the literature warn against the risks in using this indicator, related to a series of evident limitations and biases. Stimulated by this discussion, we wished to examine whether it is true that the use of impact factor should *always* be avoided in favour of citations.

We have considered a particular evaluation framework: the rating of productivity of universities in a national research system. In particular, we examine an input-output measure of productivity in which input is represented by the dimension of research staff numbers while output measures the total impact of the staff's publications, first with reference to citations and then using journal impact factor.

From the analyses conducted for the 2004-2006 triennium, it first emerges that the lists of rankings are somewhat different but are strongly correlated, although there is an element of some variability among the disciplines. However, the correlation between productivity rankings obtained by weighting output for citations and for impact factor tends to diminish with lessening of the time elapsed between the date for observing the citations and that for the publication of the article. While there is an agreement among scholars on the superiority of citations over impact factor as proxy of quality of publications for "old" articles, our findings leads to the question of which of the two proxies offers greater or lesser reliability, in the case of "young" publications. To resolve the question, a further analysis is conducted, based on their comparison to a benchmark measure of the productivity rankings obtained by considering the citations observed at "maturity" of the publication life cycles. This further analysis reveals the existence of a situation of space-time variability: the reliability of the proxy used to measure quality of individual publications is not insensitive either to scientific area for which analysis is conducted nor to the period of observation taken into consideration. Meanwhile, authoritative studies indicate that the life cycles and peaks in the citation curves for articles vary from discipline to discipline. Mathematical sciences present the most severe case of an area in which the time necessary to observe half the citations that an article will receive over its life is much greater, on average, than it is for other disciplines. These observations make it logical to hypothesize that the level of reliability of a bibliometric proxy depends on the "maturity" of the information associated with the proxy. In the case of the citation count, we can say that this becomes correspondingly less reliable in representing the quality of an article with a lessening in time passed between the publication date and the date for observation of the number of citations received. The inter-temporal comparison conducted in the final section of the examination provides confirmation: the ranking lists tend to diverge with lessening of time elapsed, and this divergence is greatest exactly in the area (Mathematics) where the life cycle of citations is significantly longer. In the context of the evaluation framework used, comparison to a sufficiently reliable bibliometric benchmark permits the demonstration that citations observed at a moment too close to the date of publication will not necessarily offer a proxy of quality that is preferable to impact factor. Yet bibliometric evaluation exercises, like all systems for outcome control, especially when designed to support the allocation of resources, should be based on observations of the most recent possible past. For evaluations over periods that are very close in time to the date of conducting the exercise, and especially in certain disciplines, the impact factor can thus be a predictor of the real impact of an article, and possibly a better one than citations.